\documentclass[aps,prd,twocolumn,superscriptaddress]{revtex4-1}

\usepackage[english]{babel}
\usepackage[utf8]{inputenc}
\usepackage{amsmath}
\usepackage{amssymb}
\usepackage{bm}
\usepackage{graphicx}

\def\be{\begin{equation}}
\def\te{\end{equation}}
\def\ee{\end{equation}}
\def\ba{\begin{eqnarray}}
\def\bea{\begin{eqnarray}}

\def\tea{\end{eqnarray}}
\def\ea{\end{eqnarray}}
\def\eea{\end{eqnarray}}

\begin{document}


\title{Primordial gravitational waves amplification from causal fluids}


\author{Nahuel Miron-Granese}
\email{nahuelmg@df.uba.ar}
\affiliation{Universidad de Buenos Aires, Facultad de Ciencias Exactas y Naturales. Departamento de F\'{i}sica, Buenos Aires C1428EGA, Argentina}

\author{Esteban Calzetta}
\email{calzetta@df.uba.ar}
\affiliation{Universidad de Buenos Aires, Facultad de Ciencias Exactas y Naturales. Departamento de F\'{i}sica, Buenos	Aires C1428EGA, Argentina}
\affiliation{CONICET—Universidad de Buenos Aires, Facultad de Ciencias Exactas y Naturales, Instituto de F\'{i}sica de Buenos Aires (IFIBA), Buenos Aires C1428EGA, Argentina}


\date{\today}

\begin{abstract}

We consider the evolution of the gravitational wave spectrum for super-Hubble modes in interaction with a relativistic fluid, which is regarded as an effective description of fluctuations in a light scalar minimally coupled field, during the earliest epoch of the radiation dominated era after the end of inflation. We obtain the initial conditions for gravitons and fluid from  quantum fluctuations at the end of inflation, and assume instantaneous reheating. We model the fluid by using relativistic causal hydrodynamics. There are two dimensionful parameters, the relaxation time $\tau$ and temperature. In particular we study the interaction between gravitational waves and the non trivial tensor (spin 2) part of the fluid energy-momentum tensor. Our main result is that the new dimensionful parameter $\tau$ introduces a new relevant scale which distinguishes two kinds of super-Hubble modes. For modes with $H^{-1}<\lambda<\tau$ the fluid-graviton interaction increases the amplitude of the  primordial gravitational wave spectrum at the electroweak transition by a factor of about $1.3$ with respect to the usual scale invariant spectrum.
\end{abstract}


\maketitle

\section{Introduction}

In this paper we shall consider the evolution of the primordial gravitational wave background during the early radiation dominated era \cite{libroweinberg2008} \cite{grishchuk1974} \cite{starobinsky1979}, from reheating after inflation up to the cosmological electroweak transition. We will use second order hydrodynamics \cite{israel1989} \cite{librorezzolla2013} as an effective theory for the matter fields, and obtain a linear theory for gravitons consistently coupled to the spin-2 component of the matter energy-momentum tensor.

Our motivation in using hydrodynamics as an effective theory comes from the highly successful description of the early evolution of the fireball created in relativistic heavy ion collisions (RHICs) by these methods, even in early stages where it is unlikely that local thermal equilibrium has been established \cite{romatschke2010} \cite{strickland2014}.

As a matter of fact, our problem bears a significant similarity to RHICs \cite{ulrichheinz2013}. Our main assumption is that among the fundamental fields there is at least one that is not conformally coupled; for simplicity we shall take this to be a light (effectively massless), minimally coupled scalar field with small coupling constants. These fields are commonly related to ``axion-like particles'' (ALPs) \cite{diana2017axion} \cite{witten2017}. Inflationary expansion brings this field to its De Sitter invariant vacuum state. However, this state is highly squeezed and its quantum fluctuations are much higher than those of the local vacuum state of adiabatic observers. Upon horizon exit, and particularly after reheating, these fluctuations lose quantum coherence and may be treated as classical particles \cite{kiefer1998} \cite{campo2008} \cite{lombardo2005} \cite{libroesteban2008} \cite{francocalzetta2011} \cite{calzettareviewcdq2012} – thus resembling the quark-gluon plasma generated in RHICs. These particles compose our “fluid”.

As we have learned from RHICs, the proper treatment of a real relativistic fluids on timescales not much larger than the fluid relaxation time requires the use of “second order” theories rather than the better known Eckart or Landau-Lifshitz formulations \cite{hiscock1983} \cite{hiscock1985} \cite{hiscock1988}; one of the main points of this paper is that this is the relevant framework for our discussion. In second order theories, the viscous part of the energy-momentum tensor, or some other equivalent variable, is considered as an independent degree of freedom following a Cattaneo-Maxwell type dynamical equation \cite{preziosi1989}. This equation, together with the Einstein equations and the relevant conservation laws, completes the fully consistent dynamics we are looking for.

During reheating and afterwards, we must distinguish between the physics of modes inside or outside the horizon. Reheating is dominated by the most out of equilibrium phenomenon in the history of our Universe, the sudden conversion of the energy-momentum of the inflaton field into radiation energy-momentum \cite{boyanovsky1996} \cite{linde1997} \cite{ramsey1997} \cite{basset2006} \cite{pearce2017}. We do not assume our scalar field is decoupled from the rest of matter, and so it partakes of this essentially nonlinear phenomenon. However, the nonlinearities are restricted by causality and therefore they are strong only within the horizon. Outside the horizon the evolution of the graviton-effective fluid system may be described accurately enough by linearized equations.

At the most basic level, a gravitational wave presents itself through an anisotropy in the rest frame of the fluid. Ideal hydrodynamics is restricted by the Pascal principle, namely, the state of the ideal fluid is defined solely by the chemical potentials associated to conserved charges (which moreover vanish for a conformal theory) and by the inverse temperature four-vector, and so it is locally isotropic on surfaces perpendicular to this vector. Moreover, for a true equilibrium state, the inverse temperature four-vector must be a (conformal) Killing vector \cite{israel1989}, and it may happen that for a given spacetime there are no such vectors. However, in that case hydrodynamics is not built on true equilibria, but only on approximated local equilibria. Any space time will allow for the construction of coordinate systems, such as Riemann normal coordinates \cite{libroesteban2008} \cite{libropetrov1969}, which look locally isotropic. Therefore, in the usual approach to hydrodynamics, temperature will be isotropic in the rest frame. The shear tensor, on the other hand, may be anisotropic, but because it is built from derivatives of a vector, it cannot have the symmetry of a spin-2 field. To account for the kind of anisotropy associated to a gravitational wave it is necessary to go beyond the usual framework by considering higher orders or else including from scratch a new spin-2 degree of freedom, as we shall do in the following. For further discussion we refer to \cite{dash2017}.

Unlike ideal and first order hydrodynamics, there is no universally accepted approach to second order hydrodynamics. However, in the linearized regime we are interested in, most formalisms converge. For simplicity, we shall adopt a 
divergence-type theory  scheme \cite{liu1972} \cite{liu1986} \cite{geroch1990} \cite{geroch1991} \cite{reula1997} \cite{esteban1998} \cite{calzetta2015} \cite{peraltaesteban2009} \cite{peralta2010divergence2} \cite{peralta2010divergence} where the conformally invariant fluid is described by a dimensionful parameter $T$ (which becomes the temperature when in equilibrium), the fluid four-velocity $u^{\mu}$ (which obeys $u_{\mu}u^{\mu}=-1$, we adopt MTW conventions) and a dimensionless, symmetric, traceless and transverse tensor $\zeta^{\mu\nu}$ ($\zeta^{\mu}_{\mu}=\zeta^{\mu\nu}u_{\nu}=0$). We scale this tensor so that in the linearized theory $\zeta_{\mu\nu}=\Pi_{\mu\nu}/\rho$, where $\Pi_{\mu\nu}$ is the viscous energy-momentum tensor and $\rho$ the energy density. 

For simplicity we shall not consider an explicit coupling of the fluid to other matter fields, the self and gauge interactions of the fluid will appear through the constitutive relations for the fluid, that is its relaxation time $\tau$ (to be discussed in Section \ref{estimatesoftau}), and its temperature. Under this approximation the equations of the model are the Einstein equations, energy-momentum conservation, and a Cattaneo-Maxwell equation for $\zeta_{\mu\nu}$ to be provided below.

In summary, we assume that at the end of reheating super-Hubble modes are in a state determined by their state at the end of inflation (namely, that reheating is so fast that no significant processing occurs during reheating itself), and then thermalize to the state determined by the dominant cosmic radiation background \cite{podolsky2006}; this thermalization is well described by linearized hydrodynamics. Moreover, at the relevant temperature scales the fluid may be regarded as composed of massless particles, whereby hydrodynamics becomes conformally invariant \cite{reula2017}. The tensor field $\zeta^{\mu\nu}$ may be decomposed into scalar (spin $0$), vector (spin $1$) and tensor (spin $2$) parts which are decoupled from each other at linear order. Our interest lies in the spin 2 part, which couples directly to the graviton field; for simplicity we shall disregard the scalar and vector sectors, and focus on the spin-2 sector alone. The spin 1 is relevant in scenarios including gauge fields, since it is related to magnetic field generation \cite{alesteban1998} \cite{aleesteban2000} \cite{aleesteban2002} \cite{aleesteban2010} \cite{aleesteban2014}.

It is well known that the spin 2 part of the matter energy-momentum tensor may seed a primordial gravitational field \cite{mollerach2004} \cite{noh2004} \cite{bartolo2004} \cite{watanabe2006} \cite{dufaux2007} \cite{price2008} \cite{matarrese2016} \cite{bethke2015} \cite{ford2017}. In the literature there are several estimates of the gravitational background created by different fields, such as the inflaton \cite{barnaby2012} \cite{figueroa2014}, the Higgs field \cite{masina2014} \cite{figueroa2007} \cite{figueroa2008}, primordial density fluctuations \cite{wands2009}, scalars and non abelian charged scalars \cite{hyde2015}, and Fermi fields \cite{figueroa2013}. In principle, the effect on the gravitational wave background may be observed through its impact on the CMB \cite{weikang2016}. The present work is closest to \cite{figueroa2015} \cite{figueroa2016} which considers the gravitational field created out of a spectator field. However, three differences stand out, namely we put the emphasis in achieving a self-consistent dynamics, including the back reaction of the gravitons on the spectator field, we incorporate the thermalization to the dominant radiation background into this picture, and we read the initial conditions for field and gravitons directly off quantum fluctuations of super-Hubble modes just before inflation ends, rather than the Starobinsky-Yokohama equation \cite{yokoyama1994}.

Let us elaborate on this last point. Under the assumption of instantaneous reheating we may obtain the initial conditions for these equations from the analysis of quantum fluctuations just before reheating. For the graviton field this is conventional, for completitude the main necessary results will be summarized below. For the effective fluid we shall treat $\zeta^{\mu\nu}$ as a stochastic Gaussian field whose self-correlation is derived from the energy-momentum self correlation of a quantum minimally coupled scalar field during inflation. Of course this is a divergent quantity, but the divergence is associated to short wavelength modes within the horizon; we shall assume a local observer will subtract the correlations corresponding to the instantaneous vacuum state (as defined by adiabatic modes), and associate the remainder with the effective fluid \cite{libroesteban2008} \cite{libroparkertoms2009} \cite{markkanen2017}.

The new dimensionful quantity $\tau$ (Eq. (\ref{definitionintegralcollision})) splits the range of super-horizon modes $k\le H$, where $H$ is Hubble’s constant during inflation, in two. For modes where $k\le \tau^{-1}$ as well, the fluid relaxation is efficient and there is no substantial effect of the fluid on the gravitons; the energy associated with the spin 2 field is just dissipated into heat. However, when $\tau^{-1}\le k\le H$ there is some amplification of the primordial gravitational spectrum due to the decay of the spin 2 part of the fluid into gravitons. This means that this mechanism may be the source of a local feature (a step) in the graviton spectrum around $k\sim\tau^{-1}\ll H$. We quantify the height of this step by solving the linearized equations from reheating up to the time of the electroweak transition, after which the primordial gravitational wave spectrum is subject to further processing \cite{libroweinberg2008}. We shall show that given appropriate values of the coupling constant (similar to some axion-like particle models) this step may fall in an observationally relevant range. This is the main result of this paper.

The paper is organized as follows. In Section \ref{causalhydrodynamics} we introduce the framework of divergence type theories from which we extract the causal hydrodynamic equations for the fluid, particularly we derive to linearized order the expression for the energy-momentum tensor and the dynamic equation for the non-equilibrium tensor. In order to deduce the system of fluid-gravitons coupled equations we gather the closure and linearized Einstein's equations in Section \ref{fluidgravitonscoupledequations}. Section \ref{initialconditions} provides the initial conditions for gravitons and non-equilibrium variable from quantum fluctuations during inflation. Section \ref{tensormodeevolution} is the main part of this paper; here we analyze the solutions of the previous system. We compute the evolution of the primordial gravitational wave spectrum for super-Hubble modes up to the electroweak transition and show that some amplification occurs for modes with $H^{-1}<\lambda<\tau$. Then we study the values of the relaxation time $\tau$ in Section \ref{estimatesoftau} from quantum field theory for a scalar field with gauge coupling constant $g$. Finally we conclude with some brief final remarks summarizing the most important results.

We add two appendices. Appendix \ref{appendixconformalinvariance} discusses the conformal invariance of fluid equations in the limit of massless particles, and Appendix \ref{appendixfouriertransform} clarifies some technical tools to calculate the Fourier transform of the noise kernel for scalar fields.

\section{Fluid dynamics from divergence-type theory}\label{causalhydrodynamics}

We assume inflation brings every non-conformally coupled matter field into its de Sitter invariant vacuum state, except the inflaton which is slowly-rolling down through its potential. We also assume an instantaneous reheating, so the universe goes from inflation to radiation domination in essentially no time \cite{podolsky2006}. When inflation ends, quantum fluctuations of non-conformally coupled fields become much higher than those of the local vacuum state of adiabatic observers. After inflation, these fluctuations enter in the nonlinear regime and decohere. It therefore becomes adequate to treat them like an effective fluid. In other words, the end of inflation sets the initial conditions for the later evolution of every field in a radiation dominated universe. The proper theoretical framework for the discussion of the further evolution is given by causal relativistic hydrodynamics. We shall follow a dissipative-type theory scheme as derived from kinetic theory for massless scalar particles obeying Bose-Einstein statistics \cite{miltonesteban2017}. To linearized order we may consider any other relevant approach, such as viscous anisotropic hydrodynamics \cite{strickland2014} \cite{florkowski2016} \cite{florkowski2015} \cite{tinti2016} \cite{rischke2016} \cite{tinti2016testing} \cite{florkowski2016non} or theories based on the so-called `Entropy Production Variational Principle' \cite{peraltaramos2013} with equivalent results.

This approach consists in formulating an ansatz for the one-particle distribution function (1pdf), parametrized by the hydrodynamic variables. Later on the hydrodynamic currents such as the particle number current and the
energy-momentum tensor are derived as moments of the parameterized 1pdf, and the corresponding equations as moments of the Boltzmann equation.

We assume a perturbed Friedmann-Robertson-Walker Universe with metric $g_{\mu\nu}=a^2(\eta)\bar{g}_{\mu\nu}$ with $a(\eta)$ the scale factor depending only on conformal time $\eta$, and $\bar{g}_{\mu\nu}=\eta_{\mu\nu}+h_{\mu\nu}$, where $\eta_{\mu\nu}$ is the Minkowsky metric (with signature $\left(-,+,+,+\right)$) and $h_{\mu\nu}$ represents the primordial gravitational waves. Upon reheating the inflaton decays into radiation which is left in a state of thermal equilibrium, namely its four-velocity $U^{\mu}_{rad}=a^{-1}U^{\mu}$ follows the conformal Killing field of the Friedmann-Robertson-Walker background ($U^{\mu}=\left(1,0,0,0\right)$), and its temperature $T_{rad}=a^{-1}\bm{T}$ decays as the inverse radius of the Universe. The spectator field, which is not decoupled from radiation, thermalizes into this state, a process which may be described by linear relaxation equations. Moreover as $p^{\mu}p_{\mu}=m^2\ll T_{rad}^2$ this theory is effectively conformally invariant. This implies the energy-momentum and non-equilibrium tensor (Eq. (\ref{nonequilibriumcurrent})) are traceless. Further the Boltzmann equation for massless particles also is conformally invariant and since the procedure of taking moments does not spoil this symmetry every conservation equation is conformally invariant as well. See Appendix \ref{appendixconformalinvariance} for details. Through conformal invariance we are able to eliminate the scale factor $a$ from all equations.

As we are interested in the equilibration process of this scalar fluid to the dominant radiation, we analyze linear perturbations around a state thermalized to the dominant radiation equilibrium state. In consequence we consider a linear deviation from a Bose-Einstein equilibrium distribution $f_0=1/\left(\exp{\left(\beta^{\mu}p_{\mu}\right)}-1\right)$ where $\beta^{\mu}=U^{\mu}/\bm{T}$. To introduce fluctuations we define the complete 1pdf as
\begin{equation}\label{1pdf}
f=\frac{1}{\exp{\left(-\displaystyle\frac{u^{\mu}p_{\mu}}{T}-\kappa\,\displaystyle\frac{\zeta^{\mu\nu}}{T^2}\,p_{\mu}p_{\nu}\right)}-1},
\end{equation}
where $u^{\mu}$, $T$ and $\zeta_{\mu\nu}$ are velocity, temperature and dimensionless non-equilibrium variable of the fluid respectively. The constant in front of $\zeta_{\mu\nu}$ is chosen so that later on we shall obtain $\zeta^{\mu\nu}=\Pi^{\mu\nu}/\rho$, where $\Pi^{\mu\nu}$ is the viscous part of the energy-momentum tensor and $\rho$ the energy density, to linear order. It has the value $\kappa=\pi^4/\left(2\,5!\,\zeta(5)\right)$ with $\zeta(n)$ the Riemann function. For the collision integral we take an Anderson-Witting linear ansatz \cite{anderson1974} \cite{anderson1974rela} \cite{takamoto2010}
\begin{equation}\label{definitionintegralcollision}
I_{col}=\frac{u_{\mu}p^{\mu}}{\tau}\left(f-f_0\right),
\end{equation}
where $\tau$ is the relaxation time of the fluid. This is an external parameter of the theory, which must be derived from consideration of the  fluid particles interactions between themselves and with radiation. We shall discuss this parameter in Section \ref{estimatesoftau}.

The idea is to decompose all fields into an (homogeneous) average and a fluctuation, and obtain linearized equations for the fluctuations. From the cosmological principle we assume the background quantities have the FRW symmetry, in particular $\zeta^{\mu\nu}$ vanishes in the background. Since our purpose is to analyze interactions between the fluid and the gravitons we consider only tensor perturbations. The linearized 1pdf reads
\begin{equation}\label{distributionlinearorder}
f\simeq f_0\left[1+(1+f_0)\,\kappa\,\frac{\zeta_{\mu\nu}}{\bm{T}^2}\, p^{\mu}p^{\nu}\right].
\end{equation}
We choose a gauge where $h_{\mu\nu}U^{\nu}=0$, due to the tensor character of perturbations also ${h^{\mu}}_{\mu}=0$. Since $\zeta^{\mu\nu}$ is transverse to the four-velocity to linear order we find $U_{\mu}\zeta^{\mu\nu}=\zeta^{\mu}_{\mu}=0$.

\subsection*{Hydrodynamic equations}

To deduce the hydrodynamic equations we define the comoving energy-momentum tensor and non-equilibrium tensor as usual \cite{liu1972} \cite{liu1986} \cite{geroch1990} \cite{geroch1991} \cite{reula1997} \cite{esteban1998} \cite{calzetta2015} \cite{peraltaesteban2009} \cite{peralta2010divergence2} \cite{peralta2010divergence} \cite{aleesteban2016} \cite{miltonesteban2017}. The fluid energy-momentum tensor  reads
\begin{equation}\label{tmunu}
\overline{T}^{\mu\nu}=\int \bar{Dp}\,p^{\mu}p^{\nu}f,
\end{equation}
and the non-equilibrium current
\begin{equation}\label{nonequilibriumcurrent}
\overline{A}^{\mu\nu\lambda}=\int \bar{Dp}\,p^{\mu}p^{\nu}p^{\lambda}f.
\end{equation}
We also need the second moment of the collision integral
\begin{equation}\label{integralcollisionsecondmoment}
\overline{I}^{\mu\nu}=\int \bar{Dp}\,p^{\mu}p^{\nu}\overline{I}_{col}.
\end{equation}
In Eqs. (\ref{tmunu})-(\ref{integralcollisionsecondmoment}) the invariant relativistic measure is
\begin{equation}\label{relativisticmeasure}
\bar{Dp}=\frac{2\prod_{\mu=0}^{4}dp_{\mu}\,\delta(p^2)}{(2\pi)^3\sqrt{-\bar{g}}}\Theta(p^0).
\end{equation}
The equations are the conservation equation for energy-momentum tensor 
\begin{equation}
{\overline{T}^{\mu\nu}}_{;\mu}=0
\end{equation}
and the closure equation for non-equilibrium current
\begin{equation}\label{closureequation}
\begin{split}
&\left({S^{\alpha}}_{\mu}{S^{\beta}}_{\nu}-\frac13S^{\alpha\beta}S_{\mu\nu}\right){\overline{A}^{\mu\nu\lambda}}_{;\lambda}=\\&=\left({S^{\alpha}}_{\mu}{S^{\beta}}_{\nu}-\frac13S^{\alpha\beta}S_{\mu\nu}\right)\overline{I}^{\mu\nu}
\end{split}
\end{equation}
where $S^{\mu}_{\nu}=\delta^{\mu}_{\nu}+U^{\mu}U_{\nu}$. 
The relevant  integrals were computed in \cite{aleesteban2016}. Here we summarize the final expressions 
\begin{equation}\label{Tmunufluido}
{{\overline{T}}^{\mu}}_{\nu}=\frac{\pi^2}{30}\bm{T}^4\left(U^{\mu}U_{\nu}+\frac13{S^{\mu}}_{\nu}+{\zeta^{\mu}}_{\nu}\right),
\end{equation}
\begin{equation}
\overline{I}^{\mu\nu}=-\frac{3\pi^2}{15}\,\frac{\zeta(6)}{\zeta(5)}\,\frac{1}{\tau}\,\bm{T}^5\zeta^{\mu\nu}
\end{equation}
and
\begin{equation}
\begin{split}
{{\overline{A}}^{\mu\nu\lambda}}&=\frac{12\zeta(5)}{\pi^2}\bm{T}^5\bigg[U^{\mu}U^{\nu}U^{\lambda}+\\&+\frac13\left(S^{\mu\nu}U^{\lambda}+S^{\mu\lambda}U^{\nu}+S^{\lambda\nu}U^{\mu}\right)\bigg]-\\&-\frac{4\,\zeta(5)}{\pi^2}\bm{T}^5\left(U^{\mu}h^{\nu\lambda}+U^{\nu}h^{\mu\lambda}+U^{\lambda}h^{\mu\nu}\right)+\\&+\frac{3\pi^2}{15}\,\frac{\zeta(6)}{\zeta(5)}\,\bm{T}^5\left(\zeta^{\mu\nu}U^{\lambda}+\zeta^{\lambda\nu}U^{\mu}+\zeta^{\mu\lambda}U^{\nu}\right).
\end{split}
\end{equation}

In order to derive the linearized equations in the following section, we consider a purely spin-2 perturbation ($\textrm{TT}$) of the energy-momentum tensor (\ref{Tmunufluido}) in mixed components and closure equation (\ref{closureequation}) to first order. These expressions are
\begin{equation}\label{tmunufluidoorden1completo}
{{{{\overline{T}}^{(1)}}^{\mu}}_{\nu}}^{\textrm{TT}}=\frac{\pi^2}{30}\bm{T}^4\,{\zeta^{\mu}}_{\nu},
\end{equation}
\begin{equation}\label{equationforzeta}
b\,{h^{\alpha\beta}}_{,0}+{\zeta^{\alpha\beta}}_{,0}+\frac{1}{\tau}\zeta^{\alpha\beta}=0.
\end{equation}
respectively, with $b=20\,\zeta^2(5)/\left(\pi^4\,\zeta(6)\right)$. If we had used a Maxwell-Juttner equilibrium distribution, we would have derived the same equation but with $b=2/9$. Note the ratio of both $b_{MJ}/b_{BE}\simeq 1.024$.

In order to relate $\tau$ with the usual transport coefficients we compute the energy-momentum tensor up to first order in $\tau$. For this purpose we may discard the interaction with gravitons taking $h_{\mu\nu}=0$. However it is need to introduce perturbations in temperature $\delta T$ and velocity $v^{\mu}$, in addition to the tensor one $\zeta^{\mu\nu}$. Then the energy-momentum tensor reads
\begin{equation}
\begin{split}
{{{{\overline{T}}^{(1)}}^{\mu}}_{\nu}}=\frac{\pi^2}{30}\bm{T}^4\,\bigg[&4\frac{\delta T}{\bm{T}}\left(U^{\mu}U_{\nu}+\frac13{S^{\mu}}_{\nu}\right)+\\&+\,\frac43\left(U^{\mu}v_{\nu}+U_{\nu}v^{\mu}\right)+{\zeta^{\mu}}_{\nu}\bigg].
\end{split}
\end{equation}
Including the velocity perturbation Eq. (\ref{equationforzeta}) becomes
\begin{equation}
{\zeta^{\alpha\beta}}_{,0}+\frac{1}{\tau}\zeta^{\alpha\beta}+b\,\sigma^{\alpha\beta}=0
\end{equation}
which implies, to first order in $\tau$, $\zeta^{\alpha\beta}=-\tau\,b\,\sigma^{\alpha\beta}$. 
In consequence by simple comparison with the usual viscous energy-momentum tensor, the well-known kinematic viscosity coefficient $\nu=b\,\tau$.

\section{Fluid-gravitons coupled equations}\label{fluidgravitonscoupledequations}
From now on we normalize $H\eta\rightarrow\eta$, $Hr\rightarrow r$, where $H$ is the Hubble constant at the moment of reheating; we also define $\eta =0$ there and  $a\left(0\right)=1$.

From the linearized Einstein's equation in mixed components we get
\begin{equation}\label{einsteinprimerorden}
{{G^{(1)}}^{\mu}}_{\nu}=\frac{1}{a^2(\eta)M_{pl}^2}\,{{{\overline{T}}^{(1)}}^{\mu}}_{\nu},
\end{equation}
with $M_{pl}$ the reduced Planck mass. We apply tensor projectors to Eq. (\ref{einsteinprimerorden}) in spatial indexes. It reads
\begin{equation}
{{{G^{(1)}}^{i}}_{j}}^{\textrm{TT}}=\frac{H^2}{2}\left[-\eta^{\rho\sigma}\partial_{\rho}\partial_{\sigma}+2\frac{a'(\eta)}{a(\eta)}\partial_{\eta}\right]h_{ij},
\end{equation}
and for ${{{\overline{T}}^{(1)i}}_{j}}^{\textrm{TT}}$ we use Eq. (\ref{tmunufluidoorden1completo}). Since $h_{ij}$ and $\zeta_{ij}$ are tensor degrees of freedom we write the following Fourier decomposition for both
\begin{equation}\label{tensordecomposition}
h_{ij}(\bm{r},\eta)=\sum_{\lambda=+,\times}\int \frac{d^3k}{(2\pi)^{3/2}}\,\epsilon_{ij}^{\lambda}(\hat{k})\,h^{\lambda}_{k}(\eta)\,e^{i\bm{k}\bm{r}},
\end{equation}
where the conformal wave number is $\bar{k}_{phys}=Hk$, $\lambda=+,\times$ indicates polarization and the polarization tensors $\epsilon_{ij}^{\lambda}(\hat{k})$ satisfy $\epsilon_{ij}^{\lambda}(\hat{k})\,\delta^{ij}=k^i\,\epsilon_{ij}^{\lambda}(\hat{k})=0$ and $\epsilon_{ij}^{\lambda}(\hat{k})\,\epsilon^{ij}_{\lambda'}(\hat{k})=\delta_{\lambda\lambda'}$. Gathering the expressions above we derive similar equations for either polarization. Dropping the $\lambda$ index in $h_k$ and $\zeta_k$, together with Eq. (\ref{equationforzeta}), we get the system of equations to linear order for $h_k$ and $\zeta_k$
\begin{equation}
\begin{cases}\label{definitionk_0}
\left[\partial^2_{\eta}+k^2+2\displaystyle\frac{a'(\eta)}{a(\eta)}\partial_{\eta}\right]h_k(\eta)=\displaystyle\frac{1}{a^2(\eta)}\,K_0\zeta_k(\eta) \\
\partial_{\eta}\zeta_k(\eta)+\displaystyle\frac{1}{\tau_0}\zeta_k(\eta)=-b\,\partial_{\eta}h_k(\eta),
\end{cases}
\end{equation}
where $K_0=\pi^2\bm{T}^4/\left(15H^2M_{pl}^2\right)$ and $\tau_0=H\tau$. In the radiation dominated era $a\left(\eta\right)=1+\eta$ and $H(\eta)=\left(1+\eta\right)^{-2}$. We change variables $\eta\rightarrow z(\eta)=k(1+\eta)$ and $h_k(z)=\chi_k(z)/z$, 
therefore
\begin{equation}\label{sistemafinal}
\begin{cases}
\partial_z^2\chi_k(z)+\chi_k(z)=\displaystyle\frac{K_0\zeta_k(z)}{z}\\
\partial_z\zeta_k(z)+\displaystyle\frac{\zeta_k(z)}{k\tau_0}=-b\,\partial_z\left(\displaystyle\frac{\chi_k(z)}{z}\right).
\end{cases}
\end{equation}
To solve our problem we need the solution of (\ref{sistemafinal}) with the appropriate initial conditions for $h_k$ and $\zeta_k$, to be discussed in next section. The magnitude of the parameter $K_0$ measures the interaction strength between the tensor degrees of freedom $\zeta_k$ and $h_k$. Using instantaneous and effective reheating $H^2\simeq g_*\frac{\pi^2}{30}\bm{T}^4/3M_{pl}^2$ where $g_*$ is the number of relativistic degrees of freedom at temperature $\bm{T}$. Since $O(10^2\textrm{ GeV})\ll \bm{T}\leq M_{pl}$, then $g_*\gtrsim 10^2$ and
\begin{equation}
K_0\simeq 6\,\frac{\rho_S}{\rho_{\gamma}}=\frac{6}{g_*}\lesssim 10^{-2}.
\end{equation}

\section{Initial conditions}\label{initialconditions}

The purpose of this section is to compute the initial conditions for $h_k$ and $\zeta_k$ at the beginning of the radiation dominated era. To do this we regard them as classical stochastic Gaussian variables with zero mean, whose self correlation matches the Hadamard propagator of the corresponding quantum operators in the Bunch-Davies vacuum at the end of inflation. 

\subsection*{Gravitons $h$}
Gravitons are tensor metric perturbations. As we have seen before there are two polarizations $h^+$ and $h^{\times}$. As it is well known \cite{fordparker1977}, the amplitude for both can be treated as massless real scalar fields. As usual, to quantize them we use decomposition (\ref{tensordecomposition}) and apply canonical quantization to the auxiliary field $\chi$ defined by $h_k(\eta)=\chi_k(\eta)/a(\eta)$. Explicitly
\begin{equation}\label{fouriercampos}
\centering
\begin{split}
&h_{ij}(\bm{r},\eta)=\frac{\chi_{ij}(\bm{r},\eta)}{a(\eta)}=\\&=\sum_{\lambda=+,\times}\int \frac{d^3k}{(2\pi)^{3/2}}\frac{\epsilon_{ij}^{\lambda}(\hat{k})}{a(\eta)}\left[\chi_k^{\lambda}(\eta)\hat{a}_{\bm{k}}+\chi_k^{\lambda\,*}(\eta)\hat{a}^{\dagger}_{\bm{-k}}\right]e^{i\bm{k}\bm{r}}.
\end{split}
\end{equation}
This field $\chi$ must be dimensionless as well as $h$. As before we obtain the same equation for both polarizations of $\chi_k$. During inflation
\begin{equation}\label{ecuacionchik}
\chi_k^{\prime\prime}+\left[k^2-\frac{a^{\prime\prime}}{a}\right]\chi_k=0.
\end{equation}
During  inflation $\eta\leq 0$ and $a(\eta)=1/(1-\eta)$. We adopt the Bunch-Davies positive frequency solution \cite{librododelson2003} of (\ref{ecuacionchik})
\begin{equation}
\chi_k^I(\eta)=\frac{H}{M_{pl}}\frac{e^{-ik\eta}}{\sqrt{2k}}\left[1+i\frac{1}{k(1-\eta)}\right].
\end{equation}

Under the scheme of instantaneous reheating our initial conditions for the evolution of Fourier components $h_k$ during radiation dominated Universe $\eta\geq 0$ are
\begin{equation}\label{condicioninicialmodo}
h_{\bm{k}}(\eta=0)=i\,\frac{H}{M_{pl}}\frac{1}{\sqrt{2k^3}}\,e_{\bm{k}}+\frac{H}{M_{pl}}\frac{1}{\sqrt{2k}}\,b_{\bm{k}}
\end{equation}
and
\begin{equation}\label{condicioninicialmodoprima}
h'_{\bm{k}}(\eta=0)=-i\frac{H}{M_{pl}}\sqrt{\frac{k}{2}}\,e_{\bm{k}},
\end{equation}
where $e_{\bm{k}}=\hat{a}_{\bm{k}}-\hat{a}^{\dagger}_{-\bm{k}}$ and $b_{\bm{k}}=\hat{a}_{\bm{k}}+\hat{a}^{\dagger}_{-\bm{k}}$. Next we assume the Landau prescription $\langle AB\rangle_{S}=1/2\,\langle 0\left|\left\{A;B\right\}\right|0\rangle$ to convert quantum expectation values into stochastic ensemble averages \cite{librolandau} \cite{hu2008}. In consequence  \begin{equation}\label{promedioestocastico1}
\begin{split}
&\langle e_{\bm{k}}e_{\bm{k}'}^*\rangle_S=\\&=\frac12 \langle0| \left\{\left(a_{\bm{k}}-a_{-\bm{k}}^{\dagger}\right);\left(a_{\bm{k}'}^{\dagger}-a_{-\bm{k}'}\right)\right\}|0\rangle_Q=\delta(\bm{k}-\bm{k}'),
\end{split}
\end{equation}
\begin{equation}\label{promedioestocastico2}
\begin{split}
&\langle b_{\bm{k}}b_{\bm{k}'}^*\rangle_S=\\&=\frac12 \langle0| \left\{\left(a_{\bm{k}}+a_{-\bm{k}}^{\dagger}\right);\left(a_{\bm{k}'}^{\dagger}+a_{-\bm{k}'}\right)\right\}|0\rangle_Q=\delta(\bm{k}-\bm{k}'),
\end{split}
\end{equation}
\begin{equation}\label{promedioestocastico3}
\begin{split}
&\langle e_{\bm{k}}b_{\bm{k}'}^*\rangle_S=\\&=\frac12 \langle0| \left\{\left(a_{\bm{k}}-a_{-\bm{k}}^{\dagger}\right);\left(a_{\bm{k}'}^{\dagger}+a_{-\bm{k}'}\right)\right\}|0\rangle_Q=0.
\end{split}
\end{equation}

For instance, initial correlation for modes outside the horizon ($k\ll 1$) at $\eta=0$ develop a scale invariant spectrum, namely
\begin{equation}
\langle h_k(\eta)h^*_{k'}(\eta')\rangle=\delta(\bm{k}-\bm{k}')\frac{H^2}{2M_{pl}^2k^3}.
\end{equation}

\subsection*{Non-equilibrium tensor $\zeta$}
This case is more complicated because there is no immediate relation between the stochastic non-equilibrium variable $\zeta$ and some canonical quantum field during inflation. Instead, we write the tensor part of the energy-momentum tensor self correlation for a minimally coupled scalar field during inflation, namely the so-called noise kernel ${{{N^{\mu}}_{\nu}}^{\rho}}_{\sigma}$. Then we match it at $\eta=0$ to the stochastic self correlation function of $\zeta$ calculated during the radiation dominated era.

The noise kernel is defined as
\begin{equation}
{{{N^{\mu}}_{\nu}}^{\rho}}_{\sigma}=\frac{1}{2}\left[\langle\left\{{T^{\mu}}_{\nu}(x),{T^{\rho}}_{\sigma}(y)\right\}\rangle-2\langle{T^{\mu}}_{\nu}(x)\rangle\langle{T^{\rho}}_{\sigma}(y)\rangle\right].
\end{equation}
Since we will take the tensor part of the noise kernel, the only possible contribution comes from the kinetic term of the energy-momentum tensor \cite{dufaux2009}. ${{{N^{\mu}}_{\nu}}^{\rho}}_{\sigma}$ was computed in \cite{guillem2010}. For the massless ($m/H\ll1$) and large scales ($r\gg1$) limit at the end of inflation ($\eta=0$), which is our case of interest, \cite{guillem2010} obtains the following result for the kinetic term contribution
\begin{equation}\label{noisekernel}
\begin{split}
&N^{ijkl}(r,\eta=0)\simeq\\&\simeq\frac{H^8}{16\pi^4\,r^4}\left[\delta^{il}\delta^{jk}-2\left(\delta^{il}\hat{r}^j\hat{r}^k+\delta^{jk}\hat{r}^i\hat{r}^l\right)+4\,\hat{r}^i\hat{r}^j\hat{r}^k\hat{r}^l\right]+\\&+(k\leftrightarrow l).
\end{split}
\end{equation}
We disregard a term which becomes constant at large separations, since it does not contribute to the tensor part.

In Fourier space we define the projector ${{{\Lambda^a}_i}^b}_j$ into tensor part (divergenceless and traceless) like
\begin{equation}\label{proyector}
{{{\Lambda^a}_i}^b}_j={M^a}_i{M^b}_j-\frac{1}{2}M^{ab}M_{ij},
\end{equation}
with
\begin{equation}
{M^a}_i={\delta^a}_i-\frac{k^ak_i}{k^2}.
\end{equation}
Recalling that $r=\left|\bm{x}-\bm{x}'\right|$, when Fourier transforming we get two different momenta for each spatial point $\bm{x}$ and $\bm{x}'$. Due to homogeneity and isotropy, the tensor part of the Fourier transformed noise kernel $N_T^{abcd}$ results
\begin{equation}\label{cuanticaproyectada}
\begin{split}
&{N_{T\,}}^{abcd}(\bm{k},\bm{k}')=\\&={{{\Lambda^a}_i}^b}_j{{{\Lambda^c}_k}^d}_l\,\,\langle\frac{1}{2}\left\{{\left.T_{\bm{k}}\right.^i}_{j}-\langle{\left.T_{\bm{k}}\right.^i}_{j}\rangle;{\left.T^*_{\bm{k}'}\right.^k}_{l}-\langle{\left.T^*_{\bm{k}'}\right.^k}_{l}\rangle\right\}\rangle=\\
&={{{\Lambda^a}_i}^b}_j{{{\Lambda^c}_k}^d}_l\,\,{{{\left.N\right.^i}_{j}}^{k}}_{l}(\bm{k},\bm{k}')=\\&=\delta\left(\bm{k}-\bm{k}'\right)\,F(k)\,\left[\Lambda^{adbc}+\Lambda^{acbd}\right],
\end{split}
\end{equation}
with
\begin{equation}\label{fkequation}
F(k)=c\,H^8\,k+O(k^2),
\end{equation}
and $c=6911/(12\,\pi^2)$ (see Appendix \ref{appendixfouriertransform}). This result provides us the quantum fluctuations from inflation. In order to match it with our fluid non-equilibrium correlation we must to subtract the local vacuum fluctuations. It is possible to show that the pathological behaviour of (\ref{noisekernel}) at short distance is caused entirely by the mentioned local vacuum fluctuations. In fact if we calculate the noise kernel using the local fourth order adiabatic vacua at time $\eta=0$ we obtain the same terms as in (\ref{noisekernel}). However computations also show that these vacuum fluctuations are only valid for small scales ($k>1$). In consequence, after the subtraction of the local vacuum, the quantum noise kernel for large scales ($k\ll 1$) is (\ref{cuanticaproyectada}).

On the other hand, we analyze the stochastic fluctuations of the fluid energy-momentum tensor in momentum space. We know that the energy-momentum tensor satisfies ${T^{\mu}}_{\nu}(\eta=0)=a^{-2}(\eta=0){{{\overline{T}}^{\,\mu}}}_{\nu}={{{\overline{T}}^{\,\mu}}}_{\nu}$. From (\ref{tmunufluidoorden1completo}) and using decomposition (\ref{tensordecomposition}), we arrive
\begin{equation}
{{{{{\overline{T}}^{\,(1)}_k}}^i}_j}^{\textrm{TT}}=\frac{\pi^2}{30}\bm{T}^4\sum_{\lambda=+,\times}\epsilon^{\lambda}_{ij}(\hat{k})\,\zeta_k^{\lambda}(\eta).
\end{equation}
Setting $\bm{k}=k\hat{z}$ and $\zeta^{\lambda}_k(\eta=0)=\zeta^{\lambda}_k$, the most general choice is
\begin{equation}
{{{\overline{T}^{(1)}_k}^i}_j}^{\textrm{TT}}=\frac{\pi^2}{30}\bm{T}^4\left(\begin{array}{ccc}
\zeta^+_k & \zeta^{\times}_k & 0 \\
\zeta^{\times}_k & -\zeta^+_k & 0 \\
0 & 0 & 0
\end{array}\right)^{ij}.
\end{equation}
The projected correlation at  time zero is
\begin{equation}\label{stochasticcorrelation}
\begin{split}
&{{{\Lambda^a}_i}^b}_j{{{\Lambda^c}_k}^d}_l\,\langle {{T^{(1)}_k}^i}_j{{T^{(1)*}_{k'}}^k}_l\rangle={{{\Lambda^a}_i}^b}_j{{{\Lambda^c}_k}^d}_l\,\langle {{{\overline{T}}^{(1)}_k}^i}_j{{{\overline{T}}^{(1)*}_{k'}}^k}_l\rangle=\\
&=\frac{\delta(\bm{k}-\bm{k}')\bm{T}^8\pi^4}{30^2}\langle\left(\begin{array}{ccc}
\zeta^{+}_k & \zeta^{\times}_k & 0 \\
\zeta^{\times}_k & -\zeta^{+}_k & 0 \\
0 & 0 & 0
\end{array}\right)^{ab}\left(\begin{array}{ccc}
\zeta^{+*}_{k'} & \zeta^{\times*}_{k'} & 0 \\
\zeta^{\times*}_{k'} & -\zeta^{+*}_{k'} & 0 \\
0 & 0 & 0
\end{array}\right)^{cd}\rangle
\end{split}
\end{equation}
Terms like $\langle {T^{i}}_{j}\rangle\langle {T^{k}}_{l}\rangle$ are zero to first order. Just like in the quantum case a $\delta$-function appears due to homogeneity.

We match stochastic and quantum tensor correlation comparing Eqs. (\ref{cuanticaproyectada}) and (\ref{stochasticcorrelation}) in the frame where $\bm{k}=k\hat{z}$ and the initial time $\eta=0$. It results
\begin{equation}\label{condicioninicialzeta}
\begin{split}
\langle \zeta^{\times}_{\bm{k}}\zeta^{\times*}_{\bm{k}'}\rangle=\langle \zeta^{+}_{\bm{k}}\zeta^{+*}_{\bm{k}'}\rangle=\delta(\bm{k}-\bm{k}')\left[\,d\,\left(\frac{H}{\bm{T}}\right)^8\,k+O(k^2)\right]
\end{split}
\end{equation}
and
\begin{equation}\label{condicioninicialzetacruzada}
\langle \zeta^{\times}_{\bm{k}}\zeta^{+*}_{\bm{k}'}\rangle=0,
\end{equation}
with $d=(30/\pi^2)^2\,c$ (crf. Eq. (\ref{fkequation})).

As we see both polarizations follow identical equations decoupled from each other. Henceforth we shall drop the polarization label.

\section{Tensor mode evolution}\label{tensormodeevolution}

To study the solutions of the system (\ref{sistemafinal}) we make a distinction between sub-horizon ($k/a\left(\eta\right)>H\left(\eta\right)$) and super-horizon ($k/a\left(\eta\right)<H\left(\eta\right)$) modes. Recalling $z=k\left(1+\eta\right)$ the former involve $z>1$ and the latter $z<1$.

Since we only concentrate in super-horizon modes, our analysis would be valid until modes re-enter in the horizon at $z=1$. Further we consider our model to be valid up to the electroweak transition, where new effects must be considered due to the change in the number of relativistic degrees of freedom.

In consequence we will analyze solutions in the limit $k\rightarrow 0$ and $\eta$ bounded by the condition $z=k(1+\eta)<1$ or by the electroweak time, whatever happens first. We only keep the dominant terms in the power series expansion for $k\ll 1$ valid for super-horizon modes until the electroweak transition.

We interpret $K_0$ (Eq. (\ref{definitionk_0})) as an interaction parameter between gravitons and tensor fluid modes. If $K_0=0$ gravitons decouple from the fluid. We determine its evolution by solving the first equation of (\ref{sistemafinal}) with the initial conditions (\ref{condicioninicialmodo})-(\ref{promedioestocastico3}). The dominant terms in the limit $k\ll 1$ are
\begin{equation}\label{solucionsink0}
h_{\bm{k}}(\eta)=i\,\frac{H}{M_{pl}}\frac{1}{\sqrt{2k^3}}\,e_{\bm{k}}+\frac{H}{M_{pl}}\frac{1}{\sqrt{2k}}\,b_{\bm{k}}+O\left(\sqrt{k}\,\right).
\end{equation}
So
\begin{equation}\label{correlacionsink0}
\langle h_{\bm{k}}(\eta)h_{\bm{k}'}^*(\eta)\rangle=\delta(\bm{k}-\bm{k}')\left[\frac{H^2}{2M_{pl}^2k^3}+\frac{H^2}{2M_{pl}^2k}+\dots\right]
\end{equation}

Neglecting the second term in (\ref{correlacionsink0}) we obtain the so-called scale invariant spectrum, $\langle h_{\bm{k}}(\eta)h_{\bm{k}'}^*(\eta)\rangle\propto \delta(\bm{k}-\bm{k}')/k^3$.

In the general case with $K_0\neq0$ it is enough to consider the two limiting cases of (\ref{sistemafinal}), namely $k\tau_0\ll 1$ and $k\tau_0\gg 1$. Hereafter we assume $1/\tau\ll H$; we shall discuss in the Section \ref{estimatesoftau} whether this is a realistic hypothesis.

We solve the system (\ref{sistemafinal}) with initial conditions (\ref{condicioninicialmodo})-(\ref{promedioestocastico3}) for gravitons and (\ref{condicioninicialzeta})-(\ref{condicioninicialzetacruzada}) for tensor fluid modes.

When $k\tau_0\ll 1$ ($k\ll 1/\tau\ll H$ in unnormalized units) the fluid modes decay before they can interact meaningfully with gravitons. For these modes with very large wavelengths we recover to leading order the usual scale invariant spectrum, namely the first term in Eq. (\ref{correlacionsink0}).

The most interesting case is when $k\tau_0\gg 1$. It means $1/\tau\ll k\ll H$ and enables us to neglect the term $\zeta_k/(k\tau_0)$ in equations (\ref{sistemafinal}). The system takes the form
\begin{equation}
\begin{cases}
\partial_z^2\chi_k(z)+\chi_k(z)=\displaystyle\frac{K_0\zeta_k(z)}{z}\\
\partial_z\zeta_k(z)=-b\,\partial_z\left(\displaystyle\frac{\chi_k(z)}{z}\right).
\end{cases}
\end{equation}
Then,
\begin{equation}\label{ecuacionzeta}
\zeta_k(z)=-b\,h_k(z)+C_{\bm{k}},
\end{equation}
$C_{\bm{k}}$ will be set by matching the quantum noise kernel spectrum to the correlation $\langle\zeta_k(\eta)\zeta^*_{k'}(\eta)\rangle$ at initial time $\eta=0$. We assume null cross correlation $\langle\zeta_kh_{k'}^*\rangle=0$, because both variables have different physical origin. In consequence
\begin{equation}
\langle C_{\bm{k}} C^*_{\bm{k}'}\rangle=\left.\langle\zeta_{\bm{k}}\zeta_{\bm{k}'}^*\rangle\right|_{\eta=0}+b^2\left.\langle h_{\bm{k}}h_{\bm{k}'}^*\rangle\right|_{\eta=0}.
\end{equation}
Using $\langle\zeta_kh_{k'}^*\rangle=0$ explicitly, we get
\begin{equation}
\left.\langle C_{\bm{k}}h_{\bm{k}'}^*\rangle\right|_{\eta=0}=\left.\langle h_{\bm{k}}C_{\bm{k}'}^*\rangle\right|_{\eta=0}=b\left.\langle h_{\bm{k}}h_{\bm{k}'}^*\rangle\right|_{\eta=0},
\end{equation}
so, considering initial conditions (\ref{condicioninicialmodo})-(\ref{promedioestocastico3}) and (\ref{condicioninicialzeta})-(\ref{condicioninicialzetacruzada}) we derive
\begin{equation}
\begin{split}
&\langle C_{\bm{k}} C^*_{\bm{k}'}\rangle=\\&=\delta(\bm{k}-\bm{k}')\left[\,d\,\left(\frac{H}{\bm{T}}\right)^8\,k+b^2\frac{H^2}{2M_{pl}^2k^3}+b^2\frac{H^2}{2M_{pl}^2k}\right],
\end{split}
\end{equation}
and
\begin{equation}
\begin{split}
&\left.\langle C_{\bm{k}}h_{\bm{k}'}^*\rangle\right|_{\eta=0}=\left.\langle h_{\bm{k}}C_{\bm{k}'}^*\rangle\right|_{\eta=0}=\\&=\delta(\bm{k}-\bm{k}')\left[b\,\frac{H^2}{2M_{pl}^2k^3}+b\,\frac{H^2}{2M_{pl}^2k}\right].
\end{split}
\end{equation}

The equation for $\chi_k(z)$ reads
\begin{equation}
\partial_z^2\chi_k(z)+\chi_k(z)+K_0\,b\,\frac{\chi_k(z)}{z^2}=K_0\frac{C_{\bm{k}}}{z}.
\end{equation}
Let $\chi_k=\sqrt{z}\,\psi_k$ and so $h_k=\psi_k/\sqrt{z}$, therefore
\begin{equation}
\begin{split}
&z^2\psi''_k(z)+z\psi'_k(z)+\left[z^2-\left(\frac{1}{4}-b\,K_0\right)\right]\psi_k(z)=\\&=K_0C_{\bm{k}}\sqrt{z},
\end{split}
\end{equation}
whose solution is
\begin{equation}
\begin{split}
\psi_k(z)&=\overline{C}_{1\bm{k}}J_{\nu}(z)+\overline{C}_{2\bm{k}}Y_{\nu}(z)+\\&+\frac{\pi}{2}Y_{\nu}(z)\int_{z_0}^{z}\frac{J_{\nu}(z')}{z'}K_0C\sqrt{z'}\,dz'-\\&-\frac{\pi}{2}J_{\nu}(z)\int_{z_0}^{z}\frac{Y_{\nu}(z')}{z'}K_0C_{\bm{k}}\sqrt{z'}\,dz',
\end{split}
\end{equation}
where $\nu^2=1/4-b\,K_0$, and $J_{\nu}(z)$ ($j_{\nu}(z)$) and $Y_{\nu}(z)$ ($y_{\nu}(z)$) are (spherical) Bessel's functions of first and second kind respectively. The expression for $h_k(z)$ is
\begin{equation}
\begin{split}
h_k(z)=&\,C_{1\bm{k}}\,j_{\nu-1/2}(z)+C_{2\bm{k}}\,y_{\nu-1/2}(z)+ \\
&+\frac{\pi}{2}\,K_0\,C_{\bm{k}}\left[y_{\nu-1/2}(z)\int_{z_0}^{z}j_{\nu-1/2}(z')\,dz'\right.-\\&-\left.j_{\nu-1/2}(z)\int_{z_0}^{z}y_{\nu-1/2}(z')\,dz'\right],
\end{split}
\end{equation}
Our solution in the limit $k\ll1$ is
\begin{equation}\label{solutionforhk}
h_{\bm{k}}(\eta)=\,h_{\bm{k}}(0)\,G_1(\eta,\nu)+\frac{\pi}{2}\,K_0\,C_{\bm{k}}\,G_2(\eta,\nu),
\end{equation}
where
\begin{equation}
\begin{split}
&G_1(\eta,\nu)=\\&=\frac{\left(1+\eta\right)^{-\nu-1/2}\left(-1+2\nu+\left(1+\eta\right)^{2\nu}\left(1+2\nu\right)\right)}{4\nu},
\end{split}
\end{equation}
\begin{equation}
\begin{split}
&G_2(\eta,\nu)=\\&=\frac{\left(1+\eta\right)^{-\nu-1/2}}{\nu\left(4\nu^2-1\right)}\bigg[-1+2\nu-4\nu(1+\eta)^{\nu+1/2}+\\&+(1+\eta)^{2\nu}(1+2\nu)\bigg].
\end{split}
\end{equation}
Thus, equal time self correlation for gravitons reads
\begin{equation}
\begin{split}
&\langle h_{\bm{k}}(\eta)h^*_{\bm{k}'}(\eta)\rangle=\\&=\langle h_{\bm{k}}(0)h^*_{\bm{k}'}(0)\rangle\left[G_1(\eta,\nu)+b\frac{\pi}{2}K_0G_2(\eta,\nu)\right]^2+\\&+\langle\zeta_{\bm{k}}(0)\zeta_{\bm{k}'}^*(0)\rangle\left(\frac{\pi}{2}K_0\right)^2{G_2}^2(\eta,\nu).
\end{split}
\end{equation}
Let us make an ascending series expansion in $K_0$ around zero, recalling $\nu=\sqrt{1/4-b\,K_0}$, and replace the initial correlations. In that case we obtain to leading order in $k$ and $K_0$
\begin{equation}\label{espectromodificado}
\begin{split}
&\langle h_{\bm{k}}(\eta)h^*_{\bm{k}'}(\eta)\rangle=\delta(\bm{k}-\bm{k}')\frac{H^2}{2M_{pl}^2k^3}\times\\&\times\left[1+b\,K_0\left(\frac{\pi}{2}-1\right)\left(\log{(1+\eta)}-\frac{\eta}{1+\eta}\right)\right]^2.
\end{split}
\end{equation}

Our description of the spectrum evolution holds up to a certain time $\eta_{k,max}$, depending on $k$, at which either the modes re-enter in the horizon or the electroweak transition takes place. To estimate the electroweak time $\eta_{EW}$ we use the ratio of the scale factor between the end of inflation and the electroweak transition, which is $a_{EW}/a_{EOI}=T_{EOI}/T_{EW}$. The typical energy of electroweak transition is $T_{EW}\simeq 10^2 \textrm{ GeV}$ and $T_{EOI}=T_{\gamma}=\bm{T}=10^n\textrm{ GeV}$. Therefore $a_{EW}=1+\eta_{EW}=10^{n-2}$ and $\eta_{EW}\simeq 10^{n-2}$. 

On the other hand we may find the conformal time at the re-entry in the horizon $\eta_{k,re-entry}$, which depends explicitly on $k$, from the relation $\lambda_{phys}(\eta)=\lambda_ca(\eta)$. It results $\eta_{k,\textrm{re-entry}}\simeq1/k$.

\begin{figure}
\includegraphics[width=8.6cm]{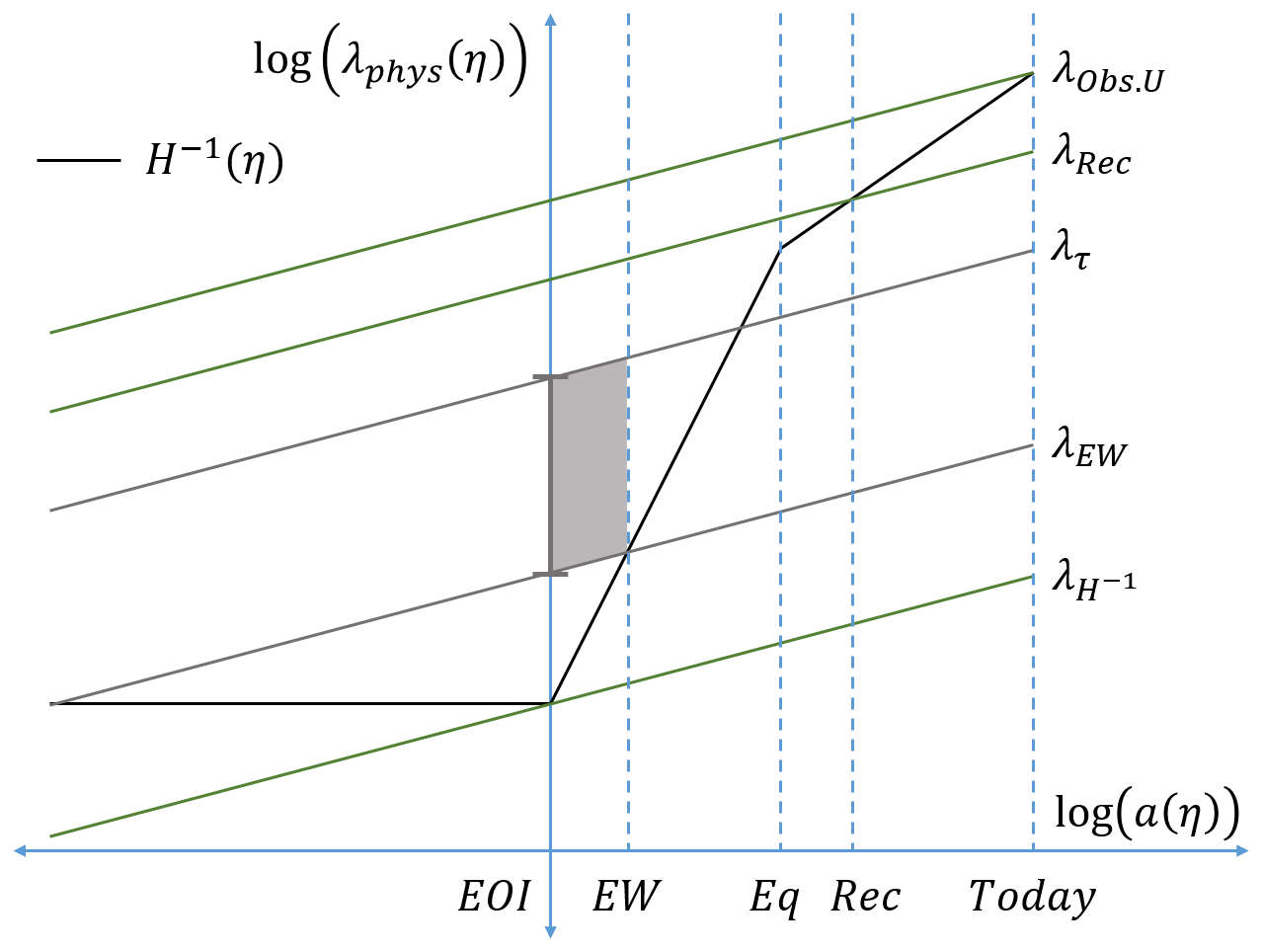}
\caption{Physical wavelength vs. scale factor.
This is a typical scheme to study evolution of perturbations during the expansion of the Universe. We show distinct events: the end of inflation ($EOI$), electroweak transition ($EW$), matter-radiation equality ($Eq$), recombination ($Rec$) and today. The Hubble radius $H^{-1}$ is represented by the black solid line and its evolution depends on the epoch of domination. $\lambda$ represents the physical wavelength of the perturbations and it scales $\lambda\propto a$. We show different wavelengths for the distinct values of the Hubble radius at the moments said. These scales are related with multipoles in the CMB correlation spectrum, for instance $l_{Rec}\sim100$. Usually $H^{-1}$ is the only relevant scale that distinguishes the evolution of perturbations between super-Hubble ($\lambda>H^{-1}$) and sub-Hubble ($\lambda<H^{-1}$) modes. We always concentrate in the former, but here it is important to note that the presence of the new dimensionful parameter $\tau$ (Eq. (\ref{definitionintegralcollision})) introduces another scale which splits the evolution of super-Hubble modes in two. First, for modes with $\lambda>\lambda_{\tau}\simeq\tau$ we recover the usual invariant spectrum. However for modes with $H^{-1}<\lambda<\lambda_{\tau}$ the fluid-graviton interaction produces an energy transfer from the fluid to gravitons and increases the amplitude of the spectrum. We are able to extend our description until the electroweak transition. Thus, shaded zone represents the modes which are amplified with respect to the usual invariant spectrum by a factor of about $1.3$ at the electroweak time according to Eq. (\ref{espectromodificadoenelectroweak}).
}
\label{fig:perturbationscheme}
\end{figure}

In Fig. \ref{fig:perturbationscheme} we show a scheme to study the evolution of physical wavelengths while the Universe expands and the horizon (Hubble radius) changes. Physical wavelengths evolve proportionally to the scale factor $a$. Modes re-enter in the horizon when $\lambda_{phys}(\eta)=H^{-1}(\eta)\rightarrow kH(\eta)/a(\eta)\simeq 1$, so the smaller the wavenumber the later its entry.

In particular at $\eta=\eta_{EW}$ one mode with comoving wavenumber $k=k_{EW}$ re-enters the horizon. Therefore the evolution of modes with $k<k_{EW}$ is bounded by $\eta_{EW}$. Conversely the time bound for modes whose $k>k_{EW}$ is $\eta_{k,\textrm{re-entry}}$.

To finish it is relevant to know what happens with $k=1/\tau$. We consider fields whose relaxation time $\tau$ produces perturbations of cosmological interest, namely perturbations whose wavelength today is at least as long as $1$ kpc. In comparison, the mode $k=k_{EW}$ has a wavelength today $\lambda_{EW,0}\lesssim 1\textrm{ pc}$, so we get $\lambda_{\tau,0}\gg\lambda_{EW,0}$, as it is shown in Fig. \ref{fig:perturbationscheme}. Therefore $1/\tau_0\ll k_{EW}$.

Summarizing, we derive the following time bounds
\begin{equation}
\begin{split}
&\eta_{k,max}=\textrm{min}_k\left\{\eta_{EW},\eta_{k,\textrm{re-entry}}\right\}=\\&=\begin{cases}
\eta_{k,\textrm{re-entry}}=\displaystyle\frac{1}{k}-1&\textrm{    if    }\,k_{EW}<k<1\\
\eta_{EW}=10^{n-2}-1&\textrm{    if    }\,\displaystyle\frac{1}{\tau_0}<k<k_{EW}.
\end{cases}
\end{split}
\end{equation}

Finally, using these bounds in Eq. (\ref{espectromodificado}) within the range of comoving unnormalized units $1/\tau<k<k_{EW}$, we obtain at $\eta=\eta_{EW}$ the gravitational wave spectrum for each polarization
\begin{equation}\label{espectromodificadoenelectroweak}
\langle h_{\bm{k}}(\eta)h^*_{\bm{k}'}(\eta)\rangle\simeq1.35\,\left(\delta(\bm{k}-\bm{k}')\frac{H^2}{2M_{pl}^2k^3}\right).
\end{equation}

\section{Estimates of $\tau$}\label{estimatesoftau}

The main goal of this section is to estimate the relaxation time $\tau$ of the field we have considered throughout the paper.

First we get a feature (step) in the spectrum at comoving wavenumber $k_{\tau}=1/\tau$ and comoving wavelength $\lambda_{\tau}=2\pi/k_{\tau}\sim\tau$. We have set $a(\eta)=1+\eta$ and $\eta=0$ at the end of inflation. For instantaneous reheating, it coincides with the onset of the radiation dominated epoch where $a_{\gamma}=a(\eta=0)=1$. The evolution of physical perturbation wavelengths from the end of inflation until today may be calculated as
\begin{equation}
\lambda_{\tau,0}=\lambda_{\tau}\,\frac{a_{0}}{a_{\gamma}}=2\pi\,\tau\,\frac{a_{0}}{a_{\gamma}},
\end{equation}
with $a_0$ the scale factor today (subscript $0$ means today). To compute the ratio $a_0/a_{\gamma}$ we consider a nearly adiabatic expansion of the Universe in which $a(\eta)\propto 1/T_{rad}$. In consequence
\begin{equation}
\frac{a_{0}}{a_{\gamma}}\simeq O(1)\,\frac{T_{\gamma}}{T_0}\simeq 10^{n+14},
\end{equation}
where $T_{\gamma}=T_{rad}(\eta=0)=10^n$ GeV is the reheating temperature. Therefore
\begin{equation}
\lambda_{\tau,0}=\lambda_{\tau}\,10^{n+14}.
\end{equation}

Recall that physical wavelengths of cosmological interest are in the range $\lambda_{0}\gtrsim 1\textrm{ kpc}$. In particular we would like to concentrate on $\lambda_{0}\gtrsim 1$ Mpc which implies $\lambda_{\tau,0}\gtrsim 1$ Mpc.

Let us consider a scalar field with a gauge coupling constant $g$. \cite{libroesteban2008} and \cite{berera1998} show that it is possible to compute the relaxation time $\tau$ in the Boltzmann equation from quantum field theory. Basically it is given by
\begin{equation}
\frac{1}{\tau}\sim\frac{\textrm{Im}\left[\Sigma\right]}{T},
\end{equation}
where $\Sigma$ is the self-energy of the field we are considering and $\textrm{Im}\left[x\right]$ takes the imaginary part of $x$. We could expand $\textrm{Im}\left[\Sigma\right]$ in Feynman diagrams and prove that the first non-null contribution appears at the two-loop order. We conclude on dimensional grounds that
\begin{equation}
\textrm{Im}\left[\Sigma\right]\sim g^4T^2=\alpha_g^2 T^2,
\end{equation}
where $\alpha_g^2=g^4$ represents the fine structure constant of this theory.

If we take the reheating temperature $T_{\gamma}\sim10^{16}-10^{15}$ GeV and 
values of $g\sim10^{-6}$ we find that $\lambda_{\tau,0}\sim10$ Mpc which lies in the range of cosmological interest. The characteristic multipole $l$ for this scale reads $l\sim\pi R_{LSS}/x\sim 10^3$, where $R_{LSS}\simeq 14$ Gpc is the distance to the last scattering surface (LSS) and $x\simeq 10$ Mpc represents the perturbation wavelength. In addition from the range of reheating temperature $T_{\gamma}\sim10^{16}-10^{15}$ GeV we consider, we estimate a tensor to scalar ratio about $r\sim 10^{-1}-10^{-5}$ respectively \cite{libroweinberg2008}.

The values of $\tau$ we are regarding here are consistent with the values for its analogous $\Gamma_{a\rightarrow\gamma\gamma}^{-1}$ (axion lifetime) in known ALP-models in the literature \cite{marsh2016} \cite{marsh2017} \cite{marsh2017_2} \cite{stadnik2017}.

We assume that the relaxation time $\tau$ and the thermalization time are of the same order and that hydrodynamics is already valid for earlier times. The validity of applying hydrodynamics in this regime has been discussed by \cite{strickland2015} \cite{romatschke2017} \cite{strickland2017} \cite{wiedemann2017} \cite{behtash2017} who argue that the hydrodynamic framework is valid at time scales shorter than the corresponding for isotropization and thermalization, driven by a novel dynamical attractor whose details vary according to the theory under consideration.

Such attractor solutions show that hydrodynamics displays a new degree of universality far-from-equilibrium regardless of the details of the initial state of the system. In fact, the approach to the dynamical attractor effectively wipes out information about the specific initial condition used for the evolution, before the true equilibrium state and consequently, thermalization, is reached. This process is described as \textit{hydrodynamization} to distinguish it from ordinary thermalization, and it has been shown by those authors that it develops on shorter time scales than thermalization.

In the context of kinetic theory and standard statistical mechanics, thermalization is understood as the development of an isotropic thermal one-particle distribution function. In some particular cases, it is possible to show that even with relative anisotropies of about 50\% the hydrodynamic description matches the full solution \cite{spalinski2017} \cite{florkowski2017}.

\section{Final Remarks}

When studying the early Universe, particularly just after inflation, it is important to include full interactions between all fields in our description. This may be a daunting challenge. In that way, we propose to treat the fields and its interactions with effective relativistic hydrodynamic theories. Nonetheless we discard ideal fluids in order to incorporate dissipative effects, as we have learned from relativistic heavy ions collisions. Further we go beyond covariant Navier-Stokes theory to avoid known causality and stability issues. Thus our main hypothesis lies in using causal hydrodynamics to obtain an adequate description of the phenomena we are interested in, specially during the very early Universe when almost all the matter fields could be described as a hot plasma.

Incorporating these causal theories to model the fields as effective fluids during the very early Universe may bring forth new effects \cite{aleesteban2016}. Throughout the paper we have analyzed a simplified case of interaction between a spectator minimally coupled scalar field and the tensor metric perturbations after inflation. Unlike ideal or Navier-Stokes hydrodynamics, this interaction may be present in any causal theory because the tensor part of the dissipative energy-momentum tensor is regarded as a new  variable with non-trivial dynamics.

Covariant Navier-Stokes equations has no proper tensor degree of freedom, in spite of the fact that the energy-momentum tensor of a quantum scalar field has such a part \cite{dufaux2007} \cite{hu2008}. Causal theories allow us to keep this component of the energy-momentum tensor and thus follow its interaction with the gravitational field. In consequence causal hydrodynamics enables the description of effects that are lost in covariant Navier-Stokes theory. Its importance would be estimated by considering the constitutive parameters. To be concrete we analyze the evolution of gravitational wave spectrum.

Usually $H^{-1}$ is the only relevant scale that distinguishes the evolution of perturbations between super-Hubble ($\lambda>H^{-1}$) and sub-Hubble ($\lambda<H^{-1}$) modes, where $\lambda$ represents the physical wavelength. We always concentrate in the former, but here it is important to note that the presence of the new dimensionful parameter $\tau$ which provides us the characteristic relaxation time of the fluid dynamics (Eq. (\ref{definitionintegralcollision})) introduces another scale which splits the evolution of super-Hubble modes in two, as it is shown in Fig. \ref{fig:perturbationscheme}. Considering the values of the parameters on previous sections we get that for modes with $\lambda>\lambda_{\tau}\simeq\tau$ we recover the usual invariant spectrum. However for modes with $H^{-1}<\lambda<\lambda_{\tau}$ the fluid-graviton interaction produces an energy transfer from the fluid to gravitons and increases the amplitude of the spectrum. We are able to extend our description until the electroweak transition. Thus, shaded zone in Fig. \ref{fig:perturbationscheme} represents the modes which are amplified with respect to the usual invariant spectrum by a factor of about $1.3$ at the electroweak time according to Eq. (\ref{espectromodificadoenelectroweak}).

Fields at extreme conditions, like highly energetic collisions or very large temperatures in the early Universe, evidence the need for new schemes of description which incorporate interactions and non-ideal processes such as dissipation and thermalization. Causal relativistic hydrodynamic theories are promising candidates to include characteristic effects of these regimes in a consistent framework.


\appendix

\section{Conformal invariance}\label{appendixconformalinvariance}

We shall show that the Boltzmann equation for massless particles is conformally invariant, and that conformal invariance is not broken by taking moments.

The Boltzmann equation in curved space is
\be 
p^{\mu}\left[\frac{\partial}{\partial x^{\mu}}+\Gamma^{\nu}_{\mu\rho}p_{\nu}\frac{\partial}{\partial p_{\rho}}\right]f=I_{col}
\te
We write $g_{\mu\nu}=a^2(\eta)\bar{g}_{\mu\nu}$. So we split the metric connection
\begin{equation}
\Gamma^{\nu}_{\mu\rho}={\bar{\Gamma}}^{\nu}_{\mu\rho}+\frac{a'}{a}\,\gamma^{\nu}_{\mu\rho},
\end{equation}
where
\begin{equation}\label{gamma}
\gamma^{\nu}_{\mu\rho}={\delta^{\nu}}_{\rho}{\delta^{0}}_{\mu}+{\delta^{\nu}}_{\mu}{\delta^{0}}_{\rho}-\bar{g}^{\nu0}\bar{g}_{\mu\rho}.
\end{equation}
We also assume that $f\left( x^{\mu},p_{\nu}\right)$ is invariant and $I_{col}=a^{-2}\overline{I}_{col}$. Thus Boltzmann equation reads
\be 
\bar{g}^{\mu\sigma}p_{\sigma}\left[\frac{\partial}{\partial x^{\mu}}+\bar{\Gamma}^{\nu}_{\mu\rho}p_{\nu}\frac{\partial}{\partial p_{\rho}}+\frac{a'}a\gamma^{\nu}_{\mu\rho}p_{\nu}\frac{\partial}{\partial p_{\rho}}\right]f=\overline{I}_{col}
\te
Conformal invariance follows if we show that
\be 
\bar{g}^{\mu\sigma}\gamma^{\nu}_{\mu\rho}\,p_{\sigma}p_{\nu}=0
\te 
for a massless theory, namely when $\bar{g}^{\mu\sigma}p_{\sigma}p_{\mu}=0$. Indeed, using (\ref{gamma}) it is straightforward to show that
\be 
\bar{g}^{\mu\sigma}\gamma^{\nu}_{\mu\rho}\,p_{\sigma}p_{\nu}=\bar{g}^{\mu\sigma}p_{\sigma}p_{\mu}{\delta^0}_{\rho}=0.
\te 
We define the covariant moments of the distribution function as
\be
A^{\mu_1,\ldots,\mu_n}=\int Dp\;p^{\mu_1}\ldots p^{\mu_n}f
\te 
where
\be 
Dp=\frac{2dp_0\prod_idp_i}{\left(2\pi\right)^3\sqrt{-g}}\;\delta\left( p^2\right)\Theta(p^0) =a^{-2}\bar{Dp},
\te
$\bar{Dp}$ is defined in Eq. (\ref{relativisticmeasure}). Then the moments transform as
\be 
A^{\mu_1,\ldots,\mu_n}=a^{-2\left( n+1\right) }\overline{A}^{\mu_1,\dots,\mu_n}
\te 
and 
\be 
I^{\mu_1,\ldots,\mu_n}=\int Dp\;p^{\mu_1}\ldots p^{\mu_n}I_{col}=a^{-2\left( n+2\right) }\overline{I}^{\mu_1,\ldots,\mu_n}
\te 
The covariant equation for the moments reads
\be
A^{\mu\mu_1,\ldots,\mu_n}_{;\mu} =I^{\mu_1,\ldots,\mu_n}
\te
and becomes
\be 
\begin{split}
&\overline{A}^{\mu\mu_1,\ldots,\mu_n}_{,\mu}+\bar{\Gamma}^{\mu}_{\mu\rho}\,\overline{A}^{\rho\mu_1,\ldots,\mu_n}+\sum_{i=1}^{n}\overline{A}^{\mu\rho\mu_1,\ldots\left(\mu_i\right)\dots,\mu_n}+\\&+\frac{a'}a\left[-2n\overline{A}^{0\mu_1,\ldots,\mu_n}+\sum_{i=1}^n\gamma^{\mu_i}_{\mu\rho}\overline{A}^{\mu\rho\mu_1,\ldots\left(\mu_i\right)\ldots,\mu_n}\right]=\\&=\overline{I}^{\mu_1,\ldots,\mu_n},
\end{split}
\te
where $\left(\mu_i\right)$ means that $\mu_i$ index is excluded. Following we need to show
\be\label{takingmomentsinvariance}
\sum_{i=1}^n\gamma^{\mu_i}_{\mu\rho}\overline{A}^{\mu\rho\mu_1,\ldots\left(\mu_i\right)\ldots,\mu_n}=2n\overline{A}^{0\mu_1,\ldots,\mu_n}
\te 
given that the moments are totally symmetric and traceless on any pair of indexes. Actually, for each term we have
\be 
\gamma^{\mu_i}_{\mu\rho}\,\overline{A}^{\mu\rho\mu_1,\ldots\left(\mu_i\right)\ldots,\mu_n}=2\overline{A}^{0\mu_1,\ldots,\mu_n}
\te 
because if $\mu_i=0$ this gives
\be
\begin{split}
&\gamma^{0}_{\mu\rho}\,\overline{A}^{\mu\rho\mu_1,\ldots\left(\mu_i\right)\ldots,\mu_n}=\\&=2\overline{A}^{00\mu_1,\ldots\left(\mu_i\right)\ldots,\mu_n}-\bar{g}^{00}{{\overline{A}^{\mu}}_{\mu}}^{\mu_1,\ldots\left(\mu_i\right)\ldots,\mu_n}=\\&=2\overline{A}^{0\mu_1,\ldots,\mu_i=0,\ldots,\mu_n}
\end{split}
\te 
and if $\mu_i=j\not =0$ then we get 
\be
\begin{split}
&\gamma^{j}_{\mu\rho}\,\overline{A}^{\mu\rho\mu_1,\ldots\left(\mu_i\right)\ldots,\mu_n}=\\&=
\overline{A}^{0j\mu_1,\ldots\left(\mu_i\right)\ldots,\mu_n}+\overline{A}^{j0\mu_1,\ldots\left(\mu_i\right)\ldots,\mu_n}-\\&-\bar{g}^{j0}{{\overline{A}^{\mu}}_{\mu}}^{\mu_1,\ldots\left(\mu_i\right)\ldots,\mu_n}=2\overline{A}^{0\mu_1,\ldots,\mu_i=j,\ldots,\mu_n},
\end{split}
\te
which ends up proving (\ref{takingmomentsinvariance}). We now show that our ansatz for the distribution function and the collision integral is consistent with conformal invariance. Indeed, we take the one-particle distribution function given in (\ref{1pdf})
\begin{equation}
f=\frac{1}{\displaystyle{\exp{\left(-\beta^{\mu}p_{\mu}-\kappa\,\zeta^{\mu\nu}\,p_{\mu}p_{\nu}/T^2\right)}}-1}.
\end{equation}
Since $p_{\mu}$ is invariant we require transformation laws which implies invariance of $\beta^{\mu}$ and $\zeta^{\mu\nu}/T^2$. Index disposition matters. From $T=\bm{T}/a$ we arrive to $\beta_{\mu}=a^2\bar{\beta}_{\mu}$, $u^{\mu}=a^{-1}\bar{u}^{\mu}$ and $\zeta_{\mu\nu}=a^2\bar{\zeta}_{\mu\nu}$. In addition as $\tau$ is a scale dimensional parameter we assume that $\tau=a\bar{\tau}$, thus
\begin{equation}
I_{col}=\frac{u^{\mu}p_{\mu}}{\tau}\left(f-f_0\right)
\end{equation}
also has the required transformation law.

\section{Tensor part of the noise kernel}\label{appendixfouriertransform}

In this appendix we clarify the calculation of tensor part of noise kernel in Fourier space. From Eq. (\ref{noisekernel}) we write
\begin{equation}
\begin{split}
&{{{{N}^{i}}_{j}}^{k}}_{l}(\bm{x},\bm{x}')=\\&=\left[r^ir^jr^kr^l\,F_1(r)+\left(\delta^{il}r^jr^k+\delta^{jk}r^ir^l\right)F_2(r)\right.+\\&+\left.\delta^{il}\delta^{jk}\,F_3(r)\right]+(k\leftrightarrow l),
\end{split}
\end{equation}
with
\begin{equation}
F_1(r)=\frac{H^8}{4\pi^4r^{8}},\;F_2(r)=-\frac{H^8}{8\pi^4r^{6}}\;\textrm{and}\;F_3(r)=\frac{H^8}{16\pi^4r^{4}}.
\end{equation}
Thus applying tensor projectors (\ref{proyector}) to (\ref{noisekernel}) in Fourier space we get
\begin{equation}
\begin{split}
&{N_{T\,}}^{abcd}(\bm{k},\bm{k}')={{{\Lambda^a}_i}^b}_j{{{\Lambda^c}_k}^d}_l\,\,{{{\left.N\right.^i}_{j}}^{k}}_{l}(\bm{k},\bm{k}')=\\&=\delta(\bm{k}-\bm{k}')\,F(k)\,\left[\Lambda^{adbc}+\Lambda^{acbd}\right],
\end{split}
\end{equation}
where
\begin{equation}
F(k)=\left[\frac{2F_1''(k)}{k^2}-\frac{2F_1'(k)}{k^3}-\frac{2F_2'(k)}{k}+F_3(k)\right].
\end{equation}
To compute Fourier transforms $F_i(k)$ we use the following relation
\begin{equation}\label{fouriertranformok}
\int\,r^{-2n}\,e^{-i\,\bm{k}\cdot\bm{r}}d^3r=\pi^{3/2}\,\frac{\Gamma(3/2-n)}{\Gamma(n)}\,\left(\frac{k^2}{4}\right)^{n-3/2},
\end{equation}
and finally it results
\begin{equation}
F(k)=\frac{6911}{12}\,\frac{H^8}{\pi^2}\,k+O(k^2).
\end{equation}

\begin{acknowledgments}
	
Work supported in part by CONICET and University of Buenos Aires. It is a pleasure to thank A. Kandus, D. López Nacir and G. Pérez-Nadal for discussions.

\end{acknowledgments}
\bibliography{paper_reference}
\end{document}